\documentclass[a4paper,12pt]{article}
\usepackage{amssymb}
\usepackage[dvips]{epsfig}
\usepackage{cite}
\newcommand{\goes}{\rightarrow}

\newcommand{\GeV}{\; \mathrm{GeV}}
\newcommand{\TeV}{\; \mathrm{TeV}}

\begin{document}
\begin{titlepage} 
\begin{flushright} 
hep-ph/9812535\\ 
\end{flushright} 

\begin{centering} 
\vspace*{1.cm} 

{\large{\bf Charge asymmetry in two-Higgs doublet models}}

\vspace*{1.cm} 
{\bf A.\ B.\ Lahanas} $^{a}$\,,\, {\bf V.\ C.\ Spanos} $^{b}$\, 
                           and\, {\bf Vasilios \ Zarikas} $^{c}$

\vspace{.7cm} 
{\it University of Athens, Physics Department,  
Nuclear and Particle Physics Section,\\  
GR--15771  Athens, Greece}\\ 
\end{centering} 

\vspace{2.cm} 
\begin{abstract}
We discuss the features of a two-Higgs doublet model exhibiting a two stage
phase transition. 
At finite temperatures electric charge violating
stationary points are developed. In conjunction with {\em CP} violation in
the Higgs or the Yukawa sector, the phase transition to the charge
conserving vacuum, generates a net charge asymmetry $\Delta Q$, in the
presence of heavy leptons, which may be well above the astrophysical bounds
put on $\Delta Q$ unless the heavy leptons are  sufficiently 
massive. This type of transition may be of relevance for supersymmetric
extensions of the Standard Model, since it shares the same features, namely
two Higgs doublets and similar {\em CP} violating sources.
\end{abstract}

\vspace{4.2cm} 
\noindent 
\rule[0.cm]{11.6cm}{.009cm} \\      
\vspace{.3cm}  
E-mail: \, $^a$ alahanas@cc.uoa.gr, $^b$ vspanos@cc.uoa.gr, $^c$ vzarikas@cc.uoa.gr 
\end{titlepage} 
\newpage 
\baselineskip=19pt 


A major problem in cosmology is the explanation of the observed baryon
asymmetry. One of the great achievements of the standard hot big bang
cosmological model is certainly nucleosynthesis, which explains the measured
abundances of the light elements in the Universe. It leads, however, to a
fine tuning problem of one parameter, the baryon to entropy ratio. It is
required that this parameter be in the range $4\times
10^{-11}<n_{B}/s<7\times 10^{-11}$, where $n_{B}$ is the net baryon density
in a commoving volume and $s$ is the entropy density.

The requirement for cosmological baryon asymmetry poses severe constraints
on the allowed particle models. In order to generate the baryon asymmetry
dynamically, as the Universe cools, the following criteria\cite{sakha},
should be satisfied:{\it \ The baryon number is not conserved, }${\it C}$%
{\it \ and }${\it CP}${\it \ symmetries are violated and
there are non-equilibrium conditions.}

The only successful attempts so far, in explaining baryogenesis at the
electroweak scale, are those based on the extensions of the Standard Model
(SM) with two-Higgs doublet models. The {\em CP} violation provided by the
Kobayashi--Maskawa matrix proves to be too small to explain cosmic baryon
asymmetry \cite{gave}. These models have a new source of {\em CP} violation,
the phase between the two vacuum expectation values of the Higgs fields 
\cite{lee,bra,bra2,wo}. Shaposhnikov \cite{shapotwo} and MacLerran \cite{mac}
first proposed that the two-Higgs model is a candidate model to explain
electroweak baryogenesis. Turok and Zadrozny subsequently analyzed how it
could work \cite{zadro}.

In supersymmetric models it is necessary to have at least two Higgs
doublets, while extra {\em CP} violating sources, occurring in other
sectors, arise naturally. In this case the magnitude of {\em CP} violation
is severely constrained by neutron's dipole moments and/or unification
assumptions, restricting the allowed parameter
space \cite{dugan,Falk,kane,tarek,barzee}.

In the two-Higgs model the desired baryogenesis at the electroweak scale is
implemented by the introduction of terms that break {\em CP} invariance
explicitly at the tree level. The necessity for introducing such terms is
the natural suppression \cite{bra2} of Higgs mediated flavour changing
neutral currents (FCNC). In fact, it is well known, that {\em CP} symmetry
is spontaneously broken in the presence of two Higgs doublets, at the
expense of introducing unnatural suppression of FCNC processes \cite
{weinberg}. On the other hand, {\em CP} violation with two Higgs doublets
and FCNC processes that are suppressed due to the heaviness of the Higgses
involved \cite{lahanas}, lead to second order phase transitions making the
baryogenesis scenario almost impossible to implement \cite{riotto}.

In the context of a two-Higgs model and in the absence of extra {\em CP}
violating terms, domains of the Universe can first tunnel to a new type of
minimum in which either of the two Higgs doublets develop no vacuum
expectation value, before evolving to the usual electroweak symmetry
breaking minimum\footnote{Land and Carlson \cite{lan}
first noticed the possibility of a two stage
phase transition. However the characteristics of it are different from this
we are discussing here.}.
In other words the phase transition occurs in two
stages. The first stage has been shown to be a first order transition, while
the subsequent is of second order \cite{zarikas}. Fluctuations of the
minimum, which we have due to the presence of the extra terms, are then
proved to lead to amplified {\em CP} violation at high 
temperatures \cite{zarikas} in the same 
spirit with Ref.~\cite{come,come2}, resulting to
reasonably large baryon asymmetry. This happens in a significant range of
the parameters that may cover the Minimal Supersymmetric Standard Model
(MSSM).

The present work is an analysis of the two-Higgs phase transition in
the sense that we fully take into account the contribution of the extra
breaking terms and do not consider them as small perturbations. The
calculation reveals a very interesting feature: The phase transition takes
place again in two stages. This happens in the entire range of the
parameters, as long as the ratio of the vev's is not fine tuned to 
unity. The explicit {\em CP} violating angles developed in this first
stage, are amplified compared to the zero temperature ones, as
required for baryogenesis. During this stage a stationary point that breaks
the $U(1)$ gauge symmetry of electromagnetism is always developed. In the
context of the two-Higgs extension of the SM the charge violating
cosmological phase does not produce a net charge asymmetry. However if the
spectrum contains heavy leptons, as we will discuss, it is possible to
create a net charge asymmetry. Such leptons, could be heavy Majorana
neutrinos. A strong motivation of having heavy neutrinos is to
explain the smallness of neutrino masses using
a see-saw mechanism \cite{Yanagida,Gell,Pilaftsis,chang}. 

The presence of a charge asymmetry can lead to dramatic effects in the
cosmological evolution due to the large strength of the electromagnetic
interactions as compared to those of gravity. Any specific model that
produces a charge asymmetry beyond a certain limit is ruled out on
cosmological grounds. Thus only models yielding small charge asymmetries are
acceptable. It should be noted that small values for the charge asymmetry
are able to produce a primordial magnetic seed field \cite{silk}.

Using weak isospin doublets both of weak
hypercharge\footnote{We follow here the notation of
Ref.~\cite{sher} in which both Higgs doublet
fields have same hypercharge.}
$Y_{{\rm weak}}=+1$ the two-Higgs scalar
potential can be written as follows \cite{sher} 
\begin{eqnarray}
V_{{\rm Higgs}} &=&\mu _{1}^{2}\Phi _{1}^{\dagger }\Phi _{1}+\mu
_{2}^{2}\Phi _{2}^{\dagger }\Phi _{2}+\lambda _{1}(\Phi _{1}^{\dagger }\Phi
_{1})^{2}+\lambda _{2}(\Phi _{2}^{\dagger }\Phi _{2})^{2}+\ \lambda
_{3}(\Phi _{1}^{\dagger }\Phi _{1})(\Phi _{2}^{\dagger }\Phi _{2})  \nonumber
\\
&&+\lambda _{4}(\Phi _{1}^{\dagger }\Phi _{2})(\Phi _{2}^{\dagger }\Phi
_{1})+{\frac{1}{2}}\lambda _{5}[(\Phi _{1}^{\dagger }\Phi _{2})^{2}+(\Phi
_{2}^{\dagger }\Phi _{1})^{2}]+V_{{\rm D}}\,,  \label{pote}
\end{eqnarray}
where $\lambda _{i}$ are real numbers and 
\begin{equation}
\Phi _{1}={\frac{1}{\sqrt{2}}}\left(
{{\phi _{1}+i\phi _{2} \atop \phi _{3}+i\phi _{4}}}\right) ,
\;\;\;\Phi _{2}={\frac{1}{\sqrt{2}}}
\left( {{\phi _{5}+i\phi _{6} \atop \phi _{7}+i\phi _{8}}}\right) \,.
\label{fields}
\end{equation}

The above potential, with the exception of $V_{\mathrm{D}}$
which we discuss
in the following, is the most general one satisfying the following discrete
symmetries: 
\begin{equation}
\Phi _{2}\rightarrow -\Phi _{2},\;\;\Phi _{1}\rightarrow \Phi
_{1},\;\;d_{R}^{i}\rightarrow -d_{R}^{i},\;\;u_{R}^{i}
\rightarrow u_{R}^{i}\,,  \label{disc}
\end{equation}
where $u_{R}^{i}$ and $d_{R}^{i}$ represent the right-handed weak
eigenstates with charges ${\frac{2}{3}}$ and $-{\frac{1}{3}}$ respectively.
All other fields involved remain intact under the above discrete symmetry.
This symmetry forces all the quarks of a given charge to interact with only
one doublet, and thus Higgs mediated flavour changing neutral currents are
absent. When this discrete symmetry is broken, during a cosmological phase
transition, it produces stable domain walls via the Kibble mechanism \cite
{kibble}. One can overcome this problem by adding terms, which break this
symmetry providing at the same time the required explicit {\em CP} violation
for baryogenesis. The most general form of that part of the potential which
breaks this discrete symmetry is 
\begin{equation}
V_{{\rm D}}=-\mu _{3}^{2}\Phi _{1}^{\dagger }\Phi _{2}+\lambda _{6}(\Phi
_{1}^{\dagger }\Phi _{1})(\Phi _{1}^{\dagger }\Phi _{2})+\lambda _{7}(\Phi
_{2}^{\dagger }\Phi _{2})(\Phi _{1}^{\dagger }\Phi _{2})\,+\,h.c.
\label{soft}
\end{equation}
This will shall call hereafter as $D-$breaking part.
The parameters $\mu _{3}$,
$\lambda _{6}$ and $\lambda _{7}$ are in general complex numbers
\begin{equation}
\mu _{3}^{2}=m_{3}^{2}\ e^{i\theta _{3}},\qquad \lambda _{6}=l_{6}\
e^{i\theta _{6}},\qquad \lambda _{7}=l_{7}\ e^{i\theta _{7}}  \label{eq5}
\end{equation}
providing explicit {\em CP} violation at the tree level.

In order to study the structure of the vacua we can perform an $SU(2)$
rotation, that sets the vev's of the fields for $\phi _{1,2,4}$ of
Eq.~(\ref{fields}) equal to zero.
Solving the system $\partial V/\partial \phi _{i}=0$
implies several different stationary points. One of them is the usual
asymmetric minimum, that respects the $U(1)$ of electromagnetism 
\begin{equation}
\Phi _{1}={\frac{1}{\sqrt{2}}}\left( {\
{{0 \atop u}}}\right) ,
\;\;\Phi _{2}={\frac{1}{\sqrt{2}}}
\left( {\ {{0 \atop ve^{i\varphi }}}}\right) .
\label{I}
\end{equation}
In Eq.~(\ref{I}) $u,v,\varphi $ are real numbers. The phase $\varphi $ is
the explicit {\em CP} violating angle that appears due to existence of the
$D-$breaking terms\footnote{The value
of the phase $\varphi$ is severely constrained by data on
neutron's dipole moment and will be taken small in the following.}.
The acceptable parameters of the model are those ensuring that the above
stationary point becomes the absolute minimum at zero temperature.

The free parameters of the model can be taken to be 
the quartic couplings $\lambda _{i}$, the ratio $\beta =u/v$ and 
the mass parameter $\mu _{3}$,
since we have the following conditions at the stationary points

\begin{eqnarray}
\mu _{1}^{2} &=&-\lambda _{1}\beta ^{2}v^{2}-\frac{1}{2}\left( \lambda
_{3}+\lambda _{4}+\lambda _{5}\right) v^{2}+m_{3}^{2}\cos \theta _{3}\ \beta
^{-1}-m_{3}^{2}\sin \theta _{3}\ \beta ^{-1}\varphi  \nonumber \\
&&-\frac{1}{4}l_{7}\cos \theta _{7}\ v^{2}\beta ^{-1}+\frac{1}{4}l_{7}\sin
\theta _{7}\ v^{2}\beta ^{-1}\varphi -\frac{3}{4}l_{6}\cos \theta _{6}\
v^{2}\beta +\frac{3}{4}l_{6}\sin \theta _{6}\ v^{2}\beta \varphi ,
\label{ex1}
\end{eqnarray}
\begin{eqnarray}
\mu _{2}^{2} &=&-\lambda _{2}v^{2}-\frac{1}{2}\left( \lambda _{3}+\lambda
_{4}+\lambda _{5}\right) \beta ^{2}v^{2}+m_{3}^{2}\cos \theta _{3}\ \beta 
\nonumber \\
&&-\frac{3}{4}l_{7}\cos \theta _{7}\ v^{2}\beta +\frac{1}{2}l_{7}\sin \theta
_{7}\ v^{2}\beta \varphi -\frac{1}{4}l_{6}\cos \theta _{6}\ v^{2}\beta ^{3}
\label{ex2}
\end{eqnarray}
and 
\begin{equation}
\varphi =-\frac{-m_{3}^{2}\sin \theta _{3}+\frac{1}{4}l_{7}\sin \theta _{7}\
v^{2}+\frac{1}{4}l_{6}\sin \theta _{6}\ v^{2}\beta ^{2}}{\lambda
_{5}v^{2}\beta -m_{3}^{2}\cos \theta _{3}+\frac{1}{4}l_{7}\cos \theta _{7}\
v^{2}+\frac{1}{4}l_{6}\cos \theta _{6}\ v^{2}\beta ^{2}}\ .  \label{eq9}
\end{equation}
In deriving Eqs.~(\ref{ex1}) and (\ref{ex2}) we have made use of the fact
that the zero temperature phase is small $\left| \varphi \right| \ll 1,$ as
discussed previously. The value of $\varphi $ can become very small either
by taking the phases $\theta _{3},\theta _{6},\theta _{7}$ to be small or by
choosing the quartic coupling $\lambda _{5}$ to be large enough. One should
choose the free parameters ensuring that the
stationary point of Eq.~(\ref{I})
is indeed a minimum and the potential is bounded from below.

Scanning the whole parameter space would be too time consuming.
Therefore in
our analysis we consider values of the quartic coupling constants such that
the ratios $\frac{\left| \lambda _{i}\right| }{g^{2}}$,
with $i=1,2,...,5$,
vary in the range from 0.01 to 1.00,
where $g$ is the $SU(2)_{L}$ coupling
constant. The set of the parameters displayed in Table~\ref{param} is a
sample which respects the experimental limits on the Higgs masses, the
condition that the potential is bounded from below, and that there is an
absolute $U(1)_{em}$ minimum at zero temperature. 
The couplings $\lambda_{6} $ and $\lambda _{7}$ should be taken 
sufficiently large to respect the
experimental limits from Higgs searches.

The cosmological phase transition of the model under
consideration can be studied using the finite temperature effective
potential. A handy feature of two-Higgs doublet models is that they can give
a stronger first order phase transition and 
thinner bubble walls \cite{cline,moreno,laine}.

In order to get a complete picture of the transition it is necessary to
explore the full range of the potential. At $T=0$ the right vacuum is chosen
to be the absolute minimum after symmetry breaking takes place. For a model
with two complex doublets this is realized by selecting the two neutral
components to develop non-vanishing vacuum expectation values. In other
words the minimum rests in a specified plane. However this by no means
ensures that during the phase transition the absolute minimum remains in the
same plane, since the field space is effectively five-dimensional. What is
usually assumed is that the minimum remains in the same plane, as in the
zero temperature case. However this picture may not be correct, leading to
misleading conclusions.

In our study we include the one-loop radiative corrections
to the tree level potential, using
the temperature corrected fermion and gauge boson masses\footnote{%
For simplicity we ignore the contribution of the scalar bosons. Their
presence makes the linear in temperature term smaller and the first order
transition weaker. However the main results of the paper regarding the
charge breaking intermediate phase, remain the same irrespectively of the
specific form of the effective potential.}. Expanding the loop integrals we
get a cubic in the scalar fields term, which stems from the Matsubara zero
mode\cite{dolan}. It is beyond the scope of this paper to discuss well known
problems associated with the validity of the perturbation and the high
temperature expansion. Thus we employ the simplified potential 
\begin{equation}
V(T)=V_{{\rm Higgs}}+{\frac{1}{8}}\,[\,\sum_{i}(M_{A}^{2})_{i}\,+\,2%
\,m_{t}^{2}\,]\,T^{2}-{\frac{1}{4\pi }}\,\sum_{i}(M_{A}^{2})_{i}^{3/2}\,\,T%
\,,  \label{vbeta}
\end{equation}
where $m_{t}^{2}=h_{t}^{2}\sum\limits_{i=5}^{8}\phi _{i}^{2}$, is
the field dependent top
quark mass and $(M_{A}^{2})_{i}$ are the
corresponding eigenvalues of the gauge boson mass
matrix 
\begin{equation}
(M_{A}^{2})^{ab}=g^{2}\sum_{k=1}^{2}\Phi _{k}^{\dagger }\,
T^{a}T^{b}\,\Phi_{k}\, .  \label{lie}
\end{equation}
The Lie algebra matrices appeared in Eq.~(\ref{lie}) are defined as follows 
\begin{eqnarray}
T^{a} &=&\sigma ^{a}\hbox{~~~for~~~}a=1,2,3\,,  \nonumber \\
T^{a} &=&tI\hbox{~~~for~~~}a=4\,,
\end{eqnarray}
with $t=g^{\prime }/g$ and $g$ ($g^{\prime }$) are the $SU(2)_{L}$ 
($U(1)_{Y} $) coupling constants.

In order to find the location of the stationary points of the potential we
adopt the following procedure: At any given temperature and for given set of
couplings we first check the shape of the potential in every two dimensional
plane $\left( \phi_{i},\,\phi _{j}\right)$ which is 
specified by $\phi_{k}=0 $, $k \neq i,j$, in order to locate 
all possible stationary points.
These can be either extrema or saddle points. Their nature is identified by
simply looking at the signs of the eigenvalues of the second derivative of
the potential.

What the analysis reveals is stated in the following:

\begin{itemize}
\item  When $\beta =1$ and at very high temperatures the symmetry is
unbroken, as expected. As the temperature drops, the first asymmetric
minimum develops. This is charge preserving, see Eq.~(\ref{I}), and lies on
the plane $\phi _{3}=\phi _{7},$ $\phi _{8}\neq 0$ and $\phi _{i}=0$. There
is a barrier between this minimum and the symmetric one, which below a
critical temperature lies higher,
so we have a first order phase transition\footnote{However
this does not guarantee bubble formation 
with kink-like walls \cite{goldenfeld}.}. 
In this case everything looks conventional. However we
should point out that the situation of having $\beta =1$ is unnatural in the
sense that a fine tuning of the parameters is required in order for this to
be realized.

\item  For $\beta <1$\thinspace the picture alters
considerably\footnote{The case
$\beta >1$ is identical to $\beta <1$ with the interchange
$\Phi_{1} \rightleftharpoons \Phi _{2}$.}. In these cases,
as for instance the
one displayed in Table~\ref{param}, an additional stationary point which is
charge breaking appears. Depending on the values of the parameters involved
it may have appeared earlier or later than the charge conserving one. For a
qualitative discussion of the thermal evolution of the Universe it suffices
to distinguish three subcases:
\end{itemize}

\begin{enumerate}
\item[(i)]  At very high temperatures only the symmetric vacuum $V_{S}$
appears. As temperature drops another local minimum $V_{CX}$, which is
charge violating, starts developing which below a critical temperature lies
lower than the symmetric minimum ($V_{CX}<V_{S}$). During this stage bubbles
of this minimum nucleate and propagate. As the Universe further cools a
stationary point, $V_{C}$, starts formatting which is charge preserving but
lies higher than $V_{CX}$. At this stage this is a saddle point, but as the
temperature further drops by about $\Delta T\sim 100 \GeV$, $V_{C}$ moves
lower and becomes the absolute minimum. At the same time $V_{CX}$ becomes
unstable and through a second order phase transition the scalar fields roll
towards the stable minimum $V_{C}$. The kind of evolution just described
occurs for instance when $\beta <0.8$, $\lambda _{6,7}>g^{2}$,
$m_{3} \lesssim 100 \GeV$ but the bound on $m_{3}$ can be relaxed if the
parameter $\beta $ is very small. However this is not the only parameter
region where this interesting subcase is realized.

\item[(ii)]  There are also values in the parameter space for which, after
the symmetric vacuum, the charge preserving minimum $V_{C}$ is developed. At
a later stage of the evolution, a charge violating saddle point $V_{CX}$
starts forming which stays higher than $V_{C}$ during the whole evolution.
Then the final stage of the evolution in this case does not differ from the
subcase discussed previously. A region of the parameter space where this
holds is for
$0.8\lesssim \beta <1$, $\lambda _{6,7}>g^{2}$, $m_{3}>110 \GeV$,
but as in the previous case, we can find other regions too where this
subcase is realized.

\item[(iii)]  There is also a parameter range where both stationary points
exist (of course one of them is the minimum and the other is a saddle point)
with comparable depths and comparable barriers from the symmetric vacuum.
Therefore transition to these points proceeds with similar tunneling rates
leading to the nucleation of both types of bubbles for a while. It is worth
mentioning here, that the tunneling rate for the transition from the
symmetric vacuum to the aforementioned stationary points receives
contributions from paths not belonging to the plane of the two points, due
to the fact that the potential is multidimensional. These corrections, which
are given in a form of a determinant \cite{tanaka} in front of the well
known exponential law \cite{dine}, although seem to affect little, there are
cases where their contribution is significantly enhanced \cite{tetra}. In
the simple cases, it suffices to use the one-loop corrected tunneling rates 
\cite{zar}.
\end{enumerate}

In order to study the minima of the potential at finite temperature in the
eight dimensional field space, we exploit the $SU(2)$ invariance and eliminate
three out of the eight degrees of freedom. In fact by an $SU(2)$ 
rotation $R(T)=e^{i\sigma ^{\alpha }f_{\alpha }(T)}$ 
the doublets $\Phi _{1},\Phi _{2}$
can be brought into the following standard form: 
\begin{equation}
\Phi _{1}={\frac{1}{\sqrt{2}}}
\left( {{0 \atop u(T)}}\right) ,
\;\;\Phi _{2}={\frac{1}{\sqrt{2}}}
\left( {{z(T) \atop w(T)}}\right) \,,
\label{cbm}
\end{equation}
where $u(T)$ is real and $z(T),$ $w(T)$ complex. The reduction to five
component simplifies the picture a great deal. With the above
parameterization the two types of the stationary points discussed
previously, are as follows:
\renewcommand{\descriptionlabel}[1]{\hspace{\labelsep}\textrm{#1}}
\begin{description}
\item[Charge conserving:] $u(T)\neq 0,$ $z(T)=0$, 
$w(T)\equiv v(T)e^{i\varphi(T)}\neq 0$ (where $v$ is real)\, ,

\item[Charge breaking:] $u(T)=0$, $z(T)\neq 0$, 
$w(T)\equiv v^{\prime}(T)e^{i\varphi ^{\prime }(T)}\neq 0$ 
(both $z$, $w$ complex)\, .
\end{description} 

There is a degeneracy in the vacuum structure of the charge breaking
minimum, since the potential $V(T)$ in Eq.~(\ref{vbeta}) has an $SU(2)$
symmetry, if one sets $\Phi _{1}=0$. Therefore, there is a manifold of
absolute minima satisfying $\phi _{i}=0$, for $\,i=1,...,4\;$, 
and $\phi_{5}^{2}+\phi _{6}^{2}+\phi _{7}^{2}+\phi _{8}^{2}=
\left| z(T)\right|^{2}+\left| w(T)\right| ^{2}$. 
This vacuum degeneracy forbids the creation
of net baryon and charge asymmetry \cite{come,come2}. What we should stress
at this point
is the significance of the $D-$breaking terms. Their presence signals the
existence of charge breaking stationary points.

Another thing that the multidimensional analysis of the phase transition
makes clear, is that the explicit {\em CP} violating phase $\varphi (T)$, at
high temperatures is significantly enhanced for all the examined sets of
parameters, as compared to its zero temperature value. 
Recall that $\varphi(T)$ is non-vanishing, provided that the 
$D-$violating terms of Eq.~(\ref{soft}) are present. 
This is true even in the case where 
the phases $\theta_{3,6,7}$ are quite small. 
This is due to large temperature depended factors
in front of the arguments of the $D-$breaking terms \cite{zarikas}.

The development of a large $\varphi(T)$ affects the baryogenesis scenario,
since the baryon asymmetry depends linearly on $\varphi(T)$. In fact within
the context of the popular non-local mechanisms, which seem to work more
efficiently, the fermion which is reflected from the bubble wall experiences
a space dependent phase. Since right-handed fermions have different
reflection coefficients from the left-handed anti-particles a baryon
asymmetry is produced \cite{tur}, which for small velocities $v_{w}$ of the
bubble wall is given by 
\begin{equation}
{\frac{n_{B}}{s}}\approx {\frac{15}{2g_{s}\pi ^{4}}}v_{w}f^{2}
\left( {\frac{m}{T_{c}}}\right) \varphi \, ,
\end{equation}
where $f$ is a Yukawa coupling and $g_{s}$ denotes the number of spin states.

A net excess of charge density $\rho _{Q}\equiv e\ n_{Q}$, is produced
during the first charge breaking stage of the phase transition, if
Sakharov's last two conditions hold as in the baryon case. If {\em C} and 
{\em CP} are conserved, then $C$ and $CP$ conjugate reactions produce equal
amounts of opposite sign charge asymmetry resulting to zero net charge
asymmetry. As we shall see, in our case we have {\em CP} violation in charge
violating processes, resulting either 
from the $D-$breaking terms in Eq.~(\ref{soft}), which are in
general complex, or from complex Yukawa couplings. The
number density of particles and antiparticles is zero in thermal
equilibrium, since the masses of a particle and its {\em CPT} conjugate
antiparticle are equal. However this is not the case during a phase
transition.

The idea of a charged Universe was first considered by Lyttleton and Bondi 
\cite{lyttleton}. The fact that the strength of electromagnetic interaction
is much larger that the gravitational one, causes any net charge asymmetry
to have serious cosmological implications. Soon, upper limits on the
magnitude of the allowed asymmetry were posed
(for recent results see Refs.~\cite{Sengupta,orito}).
The most conservative constraint comes
from the requirement, that the gravitational interaction is larger than the
electrostatic repulsion of the net charge. 
This gives $n_{Q}/s\leq 10^{-18}$, where $n_{Q}$ is the number of 
elementary charges in excess. The entropy
density is used as a measure of the commoving volume, since from entropy
conservation we get $s\propto R^{-3}$.

More stringent bounds come from the demand that the anisotropic angular
distribution of cosmic rays, induced by the electric field of the net
charge, be lower than the observed one. This leads to $n_{Q}/s\leq 10^{-30}$.

An additional bound is furnished by the observed isotropy in the cosmic
microwave background radiation. A small charge asymmetry can cause very
large anisotropies \cite{Sengupta}, through the same mechanism of the
gravitational density perturbations. Thus, charge asymmetries induce
temperature fluctuations in the microwave background (Sachs-Wolfe effect).
In this case the constraint is: $n_{Q}/s\leq 10^{-29}$. Any specific
two-Higgs doublet model which does not respect the aforementioned bounds has
to be excluded on cosmological grounds.

Now we will shall see that within the context of the SM, no net charge
asymmetry can be produced, even if the Universe passes from a charge
violating phase. However the case is very different in the presence of heavy
leptons.

As we have already discussed the phase transition happens in two stages. The 
$U(1)_{em}$ gauge symmetry can be broken during the first one. One can
realize the violation of the $U(1)_{em}$ gauge symmetry because of the
development of non-vanishing vev's in the charged components of the Higgs
doublets. The Higgs part of Lagrangian of the two-Higgs SM model is: 
\begin{equation}
{\cal L}_{{\rm Higgs}}=
\sum_{i=1,2}(D_{\mu }\Phi _{i})^{\dagger }
(D^{\mu}\Phi _{i})+{\cal L}_{{\rm Yuk}}-V_{{\rm Higgs}}\,,
\label{higgs}
\end{equation}
where the Yukawa terms for one family of fermions are 
\begin{equation}
{\cal L}_{{\rm Yuk}}=-h_{u}\overline{Q}
\widetilde{\Phi }_{2}u_{R}-h_{d}\overline{Q}\Phi _{1}d_{R}-
h_{e}\overline{L}\Phi _{1}e_{R}\,
+\,h.c.\,,
\label{eq16}
\end{equation}
and $V_{{\rm Higgs}}$ reads from Eq.~(\ref{pote}). $Q$ and $L$ denote quark
and lepton $SU(2)_{L}$ doublets, respectively. 
The Higgs doublets in Eqs.~(\ref{higgs}) and (\ref{eq16}) are as follows 
\begin{equation}
\Phi _{1,2}=\left( 
\begin{array}{c}
\phi _{1,2}^{+} \\ 
\phi _{1,2}^{0}
\end{array}
\right) \,,\,\widetilde{\Phi }_{2}\equiv -
i\sigma ^{2}\Phi _{2}^{*}=\left(\begin{array}{c}
\phi _{2}^{0*} \\ 
-\phi _{2}^{-}
\end{array}
\right) .
\end{equation}

During the charge breaking phase only the second Higgs 
doublet $\widetilde{\Phi }_{2}$ develops non-vanishing vev's, 
as have been discussed previously.
The development of charged vev's has two major implications: (a) There is a
mixing between ``neutral'' and ``charged'' gauge eigenstates resulting to
new mass eigenstates without electric charge assignment, and (b) there are
``electric charge'' violating interactions among the gauge eigenstates.

The presence of the charged vev's alters significantly the mixing of gauge
bosons $W^{\pm },W^{(3)},B$  \footnote{The $W^{\pm}$
are the usual linear combinations
${\frac{1 }{{\sqrt{2}}}}(W^{(1)}\mp iW^{(2)})$.
During the charge breaking stage the $\pm$
labels do not correspond to any physical electric charge assignment.}.
In place of the usual mixing
between the neutral gauge eigenstates $W^{(3)},B$,
which results to the mass eigenstates of $\gamma ,Z$, there is a mixing
between the neutral and charged gauge eigenstates ($W^{(3)},B,W^{\pm }$),
yielding new mass eigenstates $V_{i},\ i=1,...,4$. These eigenstates do not
have an electric charge assignment.
The second novel feature is the existence
of charge violating interactions among gauge and Higgs bosons.

The Yukawa part of the Lagrangian, after shifting Higgses from their vev's,
involves the following bilinear terms: 
\begin{eqnarray}
{\cal L}_{{\rm Yuk}} &=&-h_{u}\bar{u}_{L}u_{R}\langle \phi _{2}^{0}\rangle
^{*}-h_{d}\bar{d}_{L}d_{R}\langle \phi _{1}^{0}\rangle -h_{e}\bar{e}%
_{L}e_{R}\langle \phi _{1}^{0}\rangle  \nonumber \\
&&+h_{u}\bar{d}_{L}u_{R}\langle \phi _{2}^{-}\rangle -h_{d}\bar{u}%
_{L}d_{R}\langle \phi _{1}^{+}\rangle -h_{e}\bar{\nu}_{L}e_{R}\langle \phi
_{1}^{+}\rangle \,+\,h.c.  \label{yuk}
\end{eqnarray}
The first three terms yield the usual mass terms for the fermions, while the
rest lead to a mixing between fermions of different electric charge. There
are no additional interactions in the Yukawa sector, except those appearing
in Eq.~(\ref{yuk}).

In the following for definiteness we shall elaborate the $\beta <1$ case.
Analogous results can be obtained for the $\beta >1$ case. 
From Eq.~(\ref{yuk}) it is seen that only $d$ and $u$ 
quarks suffer a mixing. The mixing
between $e$ and $\nu $ is absent, since $\langle \phi _{1}^{+,0}\rangle=0$.
In this case the gauge and mass eigenstates of leptons are the same.

As far as the potential is concerned there are new mass terms and
interactions, that mix charge and neutral components of Higgs doublets.
These are of two types: (a) terms that arise from the $D-$violating part of
the potential, which carry {\em CP} violating sources 
($\mu _{3,}\ \lambda_{6,7}$), and (b) terms that arise from the 
rest quartic part, which are 
{\em CP} preserving. Terms carrying {\em CP} violating phase will be proved
extremely important in producing net electric charge during the first stage
of the phase transition. 

We intend to calculate the net electric charge production,
during the $U(1)_{em}$ breaking stage of the phase transition.
Therefore we focus our
attention on {\em CP}-odd charge violating interactions, that participate in
charge violating processes. {\em CP} violating sources are the complex
phases in the Higgs potential ($D-$violating terms couplings) and in the
Yukawa terms (Yukawa couplings). Although the number density of the light
leptons is abundant in the plasma, leptonic processes, such as for instance
the decay $l\rightarrow \nu _{l}\,\nu \bar{\nu}$, yield vanishing net
electric charge, because there is always sufficient thermal energy to
activate the inverse reaction, $\nu _{l}\ \nu \bar{\nu}\rightarrow$ $l\,$.
Therefore the two processes coexist in equilibrium in the plasma, resulting
to a vanishing net electric charge. Thus we turn our attention on these
charge violating processes, in which the initial state is characterized by 
a mass which is large,
as compared to the masses of the particles produced in the final
state. In these cases only when the produced particles acquire large kinetic
energies, due to their thermal motion, the inverse reaction can be
activated. On these grounds a candidate process, within SM, to succeed in
producing non-zero net charge asymmetry, is the charge violating decay
$t \rightarrow d(s)\,\nu \bar{\nu}$.
However even in this case, the net effect
is zero since there are fast strong reactions
$t \, g \rightleftharpoons d(s) \, g$,
which keep in equilibrium the quarks, washing out any asymmetry.
Thus within the SM it is not possible to gain any net charge asymmetry.
However in the case we have heavy additional particles which are out of
equilibrium and participate in charge violating reactions, a net charge
asymmetry can be produced. Such candidate particles are heavy leptons.

As an example of how a charge asymmetry may be produced we consider a simple
and favorable model
\cite{melo,non,kalyniak,pil}, in
which the fermionic spectrum of the SM is enlarged by adding two isosinglet
neutrino fields, $n^{c}$ and $S$, per family $l$. Their mass mixing, in Weyl
basis, is given by 
\begin{equation}
-{\cal L}_{mass}=\frac{1}{2}\left( 
\begin{array}{lll}
\nu & n^{c} & S
\end{array}
\right) \left( 
\begin{array}{lll}
0 & D & 0 \\ 
D^{T} & 0 & M^{T} \\ 
0 & M & 0
\end{array}
\right) \left( 
\begin{array}{l}
\nu \\ 
n^{c} \\ S
\end{array}
\right) + h.c. 
\end{equation}
The fields $\nu$, $n^{c}$, $S$ represent 
collectively all families, $\nu_{l}\equiv \left( 
\begin{array}{lll}
\nu _{e} & \nu _{\mu } & \nu _{\tau }
\end{array}
\right) $ 
so that $D$ and $M$ are actually $3 \times 3$ mass matrices. We
assume that the elements of $D$ are much smaller than those of $M$. This
model leads \cite{melo} to three massless neutrinos 
$\nu _{iL}^{^{\prime}}$, $i=1,2,3,$ one for
each family and three Dirac heavy neutrinos 
$N_{\alpha }$ , $\alpha =5,6,7$ 
with masses $m_{N}\simeq M+O(D^{2}/M)$. The
mixing can be expressed as follows 
\begin{equation}
\nu _{lL}=\sum_{i=1,2,3}(K_{L})_{li}\nu _{iL}^{^{\prime }}+\sum_{\alpha
=4,5,6}(K_{H})_{l\alpha }N_{\alpha L}\ , 
\end{equation}
with $K_{H}\sim O(\frac{D}{M})$ and $K_{L}\sim O(1)$. We can also introduce
a small Majorana mass $\mu $ in the (3,3) entry of the mass matrix if we
want the neutrinos $\nu _{i}^{^{\prime }}$ to acquire a small mass.

The relevant interactions able to produce charge asymmetry are the Yukawa
and gauge interactions
\begin{equation}
{\cal L}_{Y}=-\left( h_{N}\right) ^{ij}\left( 
\begin{array}{ll}
\overline{\nu _{L}}\; , & \overline{l_{L}}
\end{array}
\right) _{i}\ \widetilde{\Phi }_{2}\left( n_{R}\right) _{j}+h.c. \,,
\end{equation}
where $\left( h_{N}\right) ^{ij}$ are Yukawa couplings. The $S_{l}$ field
does not couple to $\Phi _{i}$. The charged current part of the gauge
interactions is given by 
\begin{equation}
{\cal L}_{cc}=\frac{1}{2\sqrt{2}}gW^{\mu }\sum_{l=e,\mu \tau }\left[ \sum_{i}%
\overline{l}\gamma _{\mu }\left( 1-\gamma _{5}\right) \left( K_{L}\right)
_{li}\nu _{i}^{\prime }+\sum_{\alpha }\overline{l}\gamma _{\mu }\left(
1-\gamma _{5}\right) \left( K_{H}\right) _{l\alpha }N_{\alpha }\right] +h.c. 
\end{equation}

\begin{figure}[t]
\begin{center}
\epsfig{file=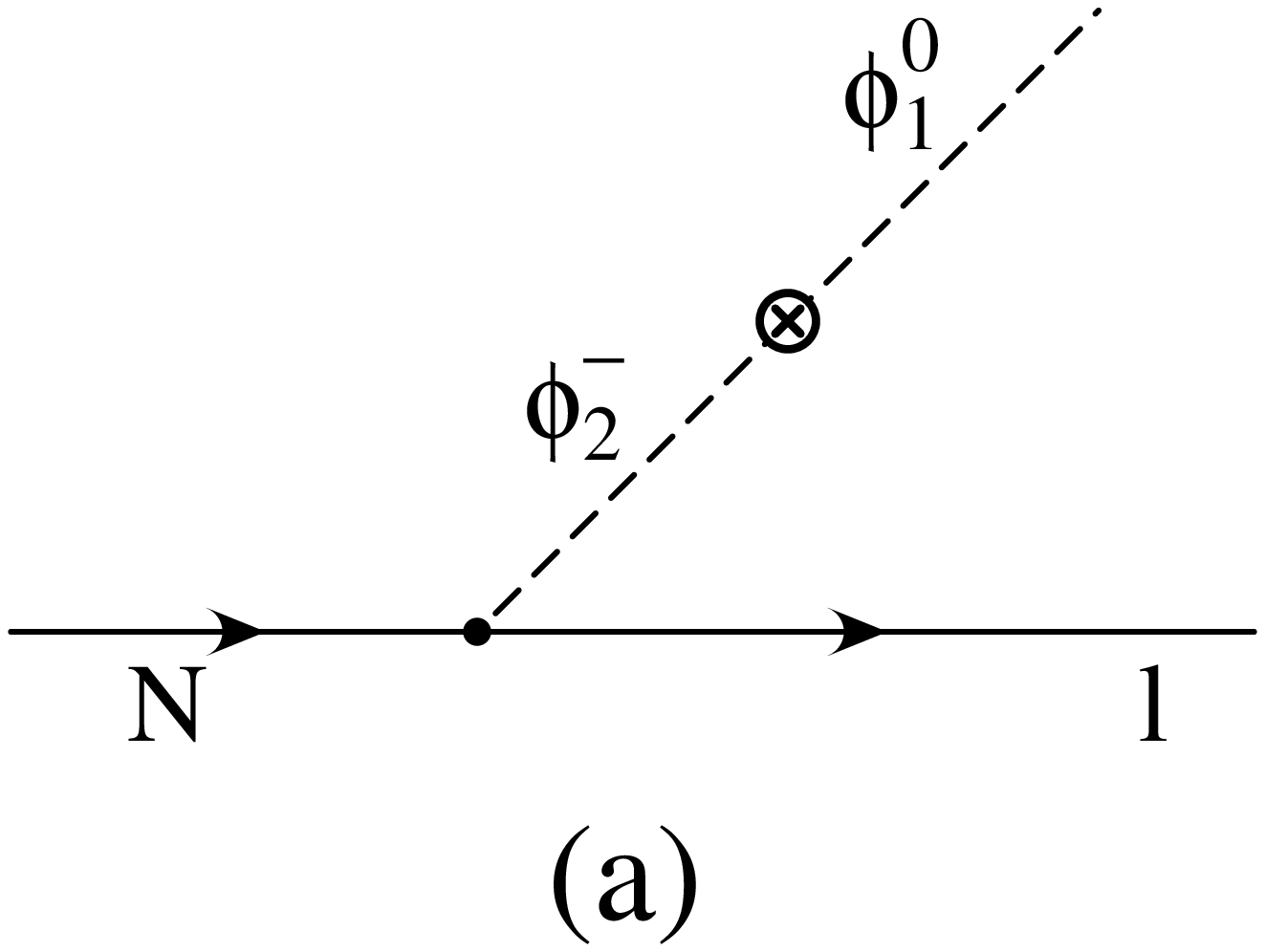,height=4.2cm,width=4.8cm} \hspace{1.5cm} %
\epsfig{file=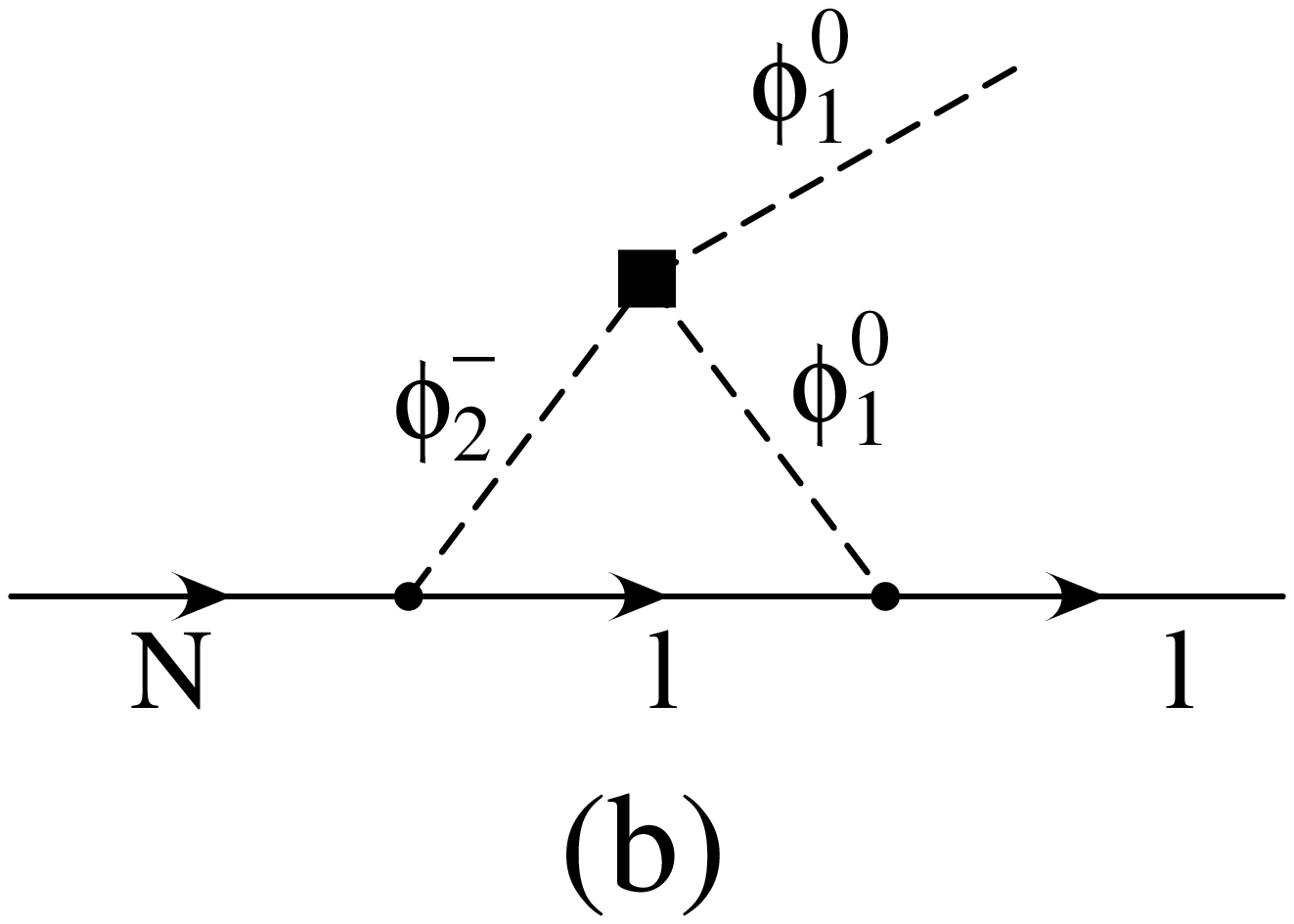,height=4.2cm,width=4.8cm} 

\vspace{1.cm}
\epsfig{file=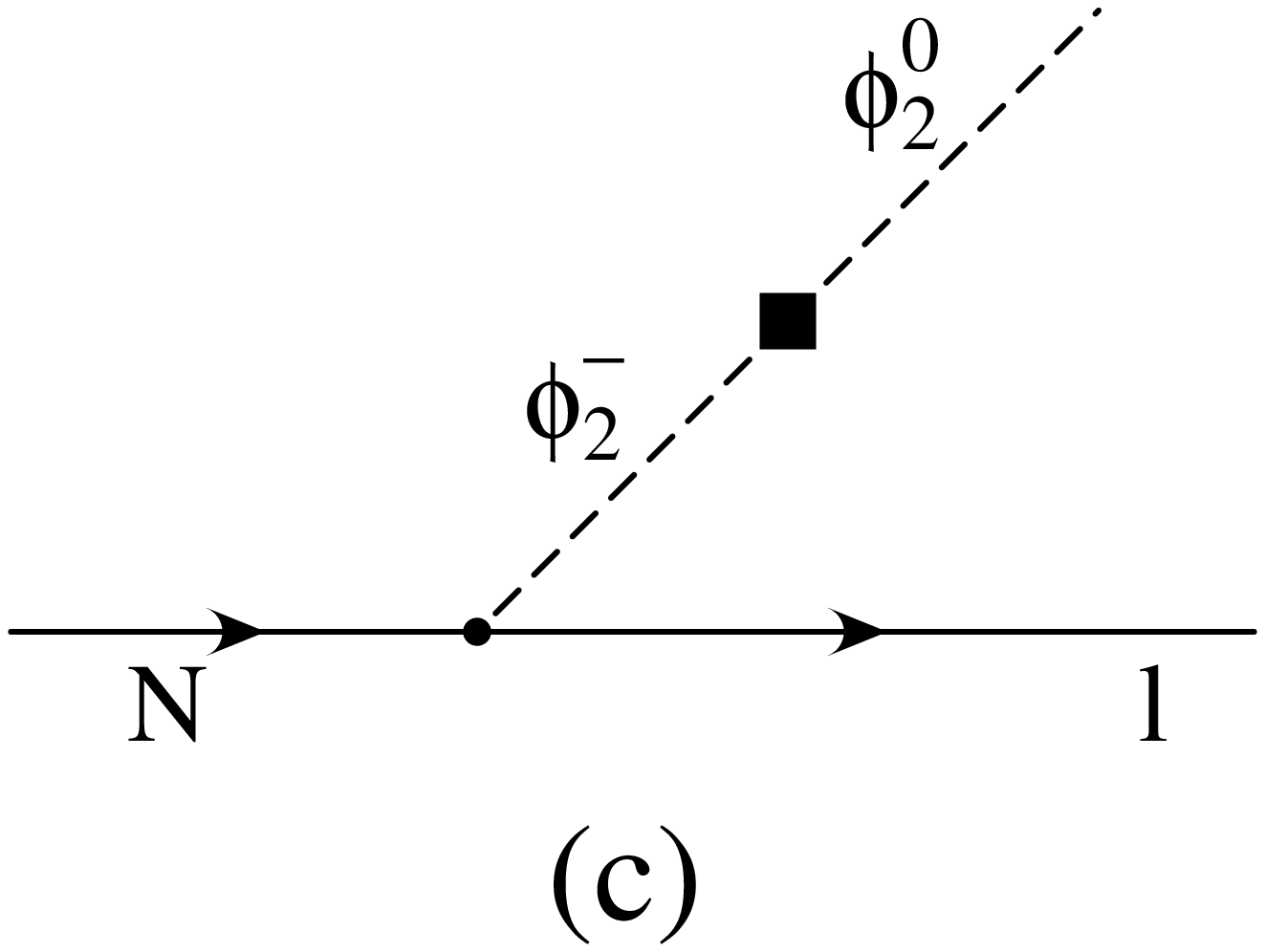,height=4.2cm,width=4.8cm} \hspace{1.5cm} %
\epsfig{file=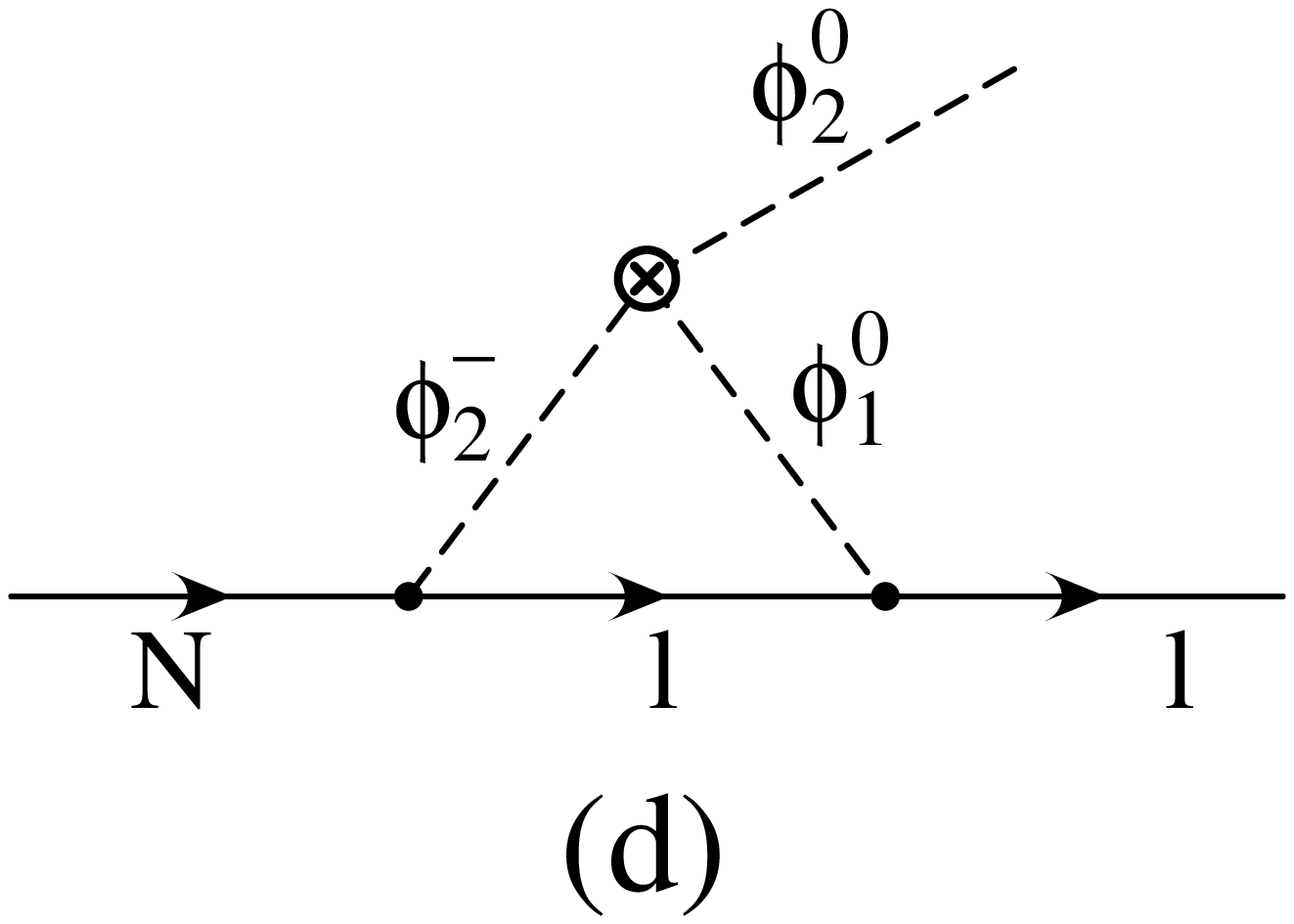,height=4.2cm,width=4.8cm} 
\begin{minipage}[t]{14.cm} 
\caption[]{Diagrams which  contribute, up to one-loop order, to 
charge violating process $N  \goes l \, \phi^0_{1,2} $.
Filled boxes (circled crosses) are used
to indicate {\em CP}-even (odd)
charge violating mixings or interactions between 
the gauge eigenstates.}
\label{fig1}
\end{minipage} 
\end{center}
\end{figure}

\begin{figure}[t]
\begin{center}
\epsfig{file=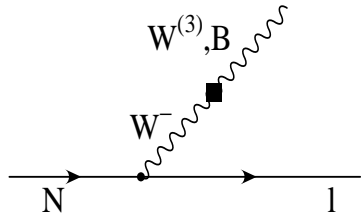,height=3.2cm,width=4.7cm} 
\begin{minipage}[t]{14.cm} 
\caption[]{ Tree level diagrams which  contribute to 
charge violating process $N  \goes l \, V $. The filled box is
as in Fig.~\ref{fig1}.}
\label{fig2}
\end{minipage} 
\end{center}
\end{figure}

Then we consider the process $N\rightarrow l\,\phi _{1,2}^{0}$,
presented in Fig.~\ref{fig1},
and for simplicity we assume one family. In order to calculate the
net electric charge production rate we compare, as in the case of baryon
asymmetry \cite{riotto,kolb}, the decay rate of process $N\rightarrow
l\,\phi _{1,2}^{0}$ and that of its {\em CP} conjugate. The difference of
these rates will be proportional to the produced average net electric
charge. In Figs.~\ref{fig1},\ref{fig2}
filled boxes (circled crosses) are used to indicate
{\em CP}-even (odd) charge violating mixings or interactions between the
gauge eigenstates. The circled crosses depend on the complex 
couplings $\mu_{3},\lambda _{6,7}$. 
All lines in Figs.~\ref{fig1},\ref{fig2}
represent gauge eigenstates,
just in order to make apparent the charge violating mixings and
interactions. These gauge eigenstates are linear combinations of the
corresponding mass eigenstates.

In Fig.~\ref{fig1}a the tree level graph does
depend on the {\em CP} violating
complex parameter $\lambda _{7}$. In the Born approximation there is no
net electric charge production, since the cross section remains invariant
under the {\em CP} transformation, even if complex couplings are involved: 
\begin{equation}
\Gamma ^{{\rm Born}}(N\rightarrow l\,\phi _{1}^{0})=\Gamma ^{{\rm Born}}(%
\overline{N}\rightarrow \overline{l}\,\phi _{1}^{0*})\,.
\end{equation}
The one-loop diagram displayed in Fig.~\ref{fig1}b
depends linearly on the Yukawa
coupling $h_{l}$, which can be complex in general,
and on $\lambda _{3}$ which is
however real. The complexity of $\lambda _{7}$ and Yukawa couplings results
to non-vanishing net electric charge production from the interference term
of the tree level graph in Fig.~1a and the one-loop graph in
Fig.~\ref{fig1}b. The
contribution of this interference term can be written as: 
\begin{equation}
\Gamma ^{{\rm interf}}(N\rightarrow l\,\phi _{1}^{0})=\lambda _{3}h_{l}\
I_{Nl}\,\left| \left\langle {\phi _{2}^{-}}\right\rangle \right| {\lambda }%
_{7}{(}h_{N})^{2}+\,h.c.\,,  \label{eq1}
\end{equation}
where $\,I_{Nl}$ is the relevant kinematic factor corresponding to this
interference term. For simplicity we have assumed that $h_{N}$ is real. The 
{\em CP} dependence of these elements is of the order of the phases
of $h_{l} $ and $\lambda _{7}$.
The {\em CP} conjugate process
$\overline{N}\rightarrow \overline{l}\,\phi _{1}^{0*}$
yields an interference
contribution: 
\begin{equation}
\Gamma ^{{\rm interf}}(\overline{N}\rightarrow \overline{l}\,\phi
_{1}^{0*})=\,\,\lambda _{3}h_{l}^{*}\ I_{Nl}^{*}\,\left| \left\langle {\phi
_{2}^{-}}\right\rangle \right| {\lambda }_{7}^{*}{(}h_{N})^{2}+\,h.c.\,,
\label{eq2}
\end{equation}
where in both Eqs.~(\ref{eq1}) and (\ref{eq2}), we have used the modulus of $%
\langle \phi _{2}^{-}\rangle $ , since spontaneous {\em CP} violating phases
cancel each other.

The net electric charge produced during the first stage of the phase
transition through the process $N\rightarrow l\,\phi _{1}^{0}$
is proportional, at the one-loop order, to the quantity 
\begin{equation}
\Gamma ^{{\rm interf}}(N\rightarrow l\,\phi _{1}^{0})-
\Gamma ^{{\rm interf}}(%
\overline{N}\rightarrow \overline{l}\,\phi _{1}^{0*})\simeq 4\,
\lambda_{3}\,\left| \left\langle {\phi _{2}^{-}}\right\rangle \right| \,
{\mathop{\rm Im}}[h_{l}{\lambda }_{7}]\ 
\mathop{\rm Im}
[\ I_{Nl}]\,h_{N}^{2}.  \label{cap1}
\end{equation}
To leading order, in the small {\em CP} violating phases, the following
approximation holds 
\begin{equation}
\,{
\mathop{\rm Im}
}[h_{l}{\lambda }_{7}]\ \simeq \left| {\lambda }_{7}\right| \left|
h_{l}\right| \left( \theta _{7}+{\arg }(h_{l})\right) \,.
\end{equation}
The average net electric charge produced via the above charge violating
process, measured in units of the electron charge, is therefore 
\begin{equation}
\Delta Q={\ \frac{1}{{\Gamma }_{N}}}\left[ \Gamma ^{{\rm interf}%
}(N\rightarrow l\,\phi _{1}^{0})-\Gamma ^{{\rm interf}}(\overline{N}%
\rightarrow \overline{l}\,\phi _{1}^{0*})\right] \,,  \label{cap2}
\end{equation}
where ${\Gamma }_{N}\simeq \frac{h_{N}^{2}}{8\pi }M$.
Using Eq.~(\ref{cap1}), Eq.~(\ref{cap2}) can take the form 
\begin{equation}
\Delta Q\simeq {\frac{32\pi \lambda _{3}}{M}}\,\,\left| \left\langle {\phi }%
_{2}^{-}\right\rangle \right| \left| {\lambda }_{7}\right| \,\,\left|
h_{l}\right| \left( \theta _{7}+{\arg }(h_{l})\right) \,{%
\mathop{\rm Im}}\left[ I_{Nl}\right] \,.  \label{capf1}
\end{equation}
For the process $N \rightarrow l \,\phi _{2}^{0}$, presented in
Fig.~\ref{fig1}c-\ref{fig1}d, we
get a similar contribution to $\Delta Q$ with $\lambda _{3}$
replaced by $\lambda _{2}$.
So the net charge density produced during the charge
violating
phase, if all the {\em CP} violating
phases are of order $10^{-4}$ as required by baryogenesis scenarios, turn
out to be 
\begin{equation}
\Delta Q\simeq \frac{10^{-8}}{M/\TeV }\,,
\end{equation}
leading to 
\begin{equation}
n_{Q}/s\simeq \frac{\Delta Q\,\ n_{{\rm N}}}{g_{*}\,n_{\gamma }}\simeq
\left( \frac{10^{-8}}{M/ \TeV }\right) 
\frac{n_{{\rm N}}}{g_{*}\,n_{\gamma }}.  
\label{eq27}
\end{equation}
For $M=1 \TeV$, $\frac{n_{{\rm N}}}{g_{*}\,n_{\gamma }}\simeq 10^{-5}$,
yielding $n_{Q}/s\simeq 10^{-13}$ which is clearly much larger than the
bounds $10^{-29}$--$10^{-30}$ quoted previously.
However the heavy neutrino density $n_{{\rm N}}$
falls rapidly with increasing $M$ resulting to a smaller charge
asymmetry. In order to find $n_{{\rm N}}$ at the electroweak
scale $T=100 \GeV$ one has to solve the appropriate Boltzmann equation.
By solving
this, we find that the bound $10^{-30}$ on $n_{Q}/s$ is saturated for a mass 
$M\approx 5 \TeV$. For larger values of $M$ the aforementioned
upper bound on $n_{Q}/s$ is always satisfied. Thus in such models the
cosmological upper limits on the charge asymmetry impose lower bounds on the
heavy neutrino masses, unless the combination of phases appearing 
in Eq.~(\ref{capf1}) is fine tuned to values 
much smaller than $10^{-4}$.

There are also other charge violating processes 
$N\rightarrow l\,V$ ($V$ denotes the neutral gauge bosons) that
can produce charge asymmetry through the interference of the tree-level graph,
presented in Fig.~\ref{fig2}, with the corresponding one-loop graphs.
This process is less significant than the one we considered above since its
rate is $\frac{M_{w}}{M}$ times smaller than the rate of the previous
interaction. However if $\theta _{7}={\arg }(h_{l})=0$ then
the process $N\rightarrow l\,V$ is
the dominant mechanism for charge asymmetry
provided $\theta _{3}$, $\theta _{6}$ are different from zero.
When the 
the Yukawa coupling $h_{N}$ of the heavy leptons
is complex, the process $N\rightarrow l\,\phi _{1,2}^{0}$ can lead to
non-zero charge asymmetry if we adopt the model of Ref.~\cite{pil} and use
a similar interference pattern with that presented in 
Ref.~\cite{fugu}.

To summarize, in this paper we discuss the phase transition of a model with
two Higgs doublets, in which {\em CP} violating terms are present in the
scalar potential. The motivation for this study originates from the fact
that the successful mechanisms for explaining baryogenesis are based on such
models. Besides this, such an extended Higgs sector resembles that of
supersymmetric (SUSY) models and hence our conclusions may be of relevance
for SUSY extensions of the SM.

Using the finite temperature one-loop corrected effective potential we 
found that the phase transition occurs in two stages. During the first stage
of the transition {\em CP} violating angles are amplified, as required by
baryogenesis, and simultaneously a charge breaking stationary point is
developed which is a minimum, in a wide range of the parameter space. This
leads to a non-vanishing charge asymmetry, in the presence of heavy leptons,
due to the appearance of {\em CP} violating sources within the $D-$breaking
terms and/or Yukawa couplings. In the context of the SM no net charge
asymmetry is produced.
However in models with heavy leptons the magnitude of the asymmetry
is found to exceed existing astrophysical bounds, constraining the mass
spectrum of heavy leptons. In particular in models with heavy neutrinos
we find that heavy neutrino masses smaller than $5 \TeV$ result to
unacceptably large charge asymmetry.

The study undertaken in this paper maybe of relevance for supersymmetric
models and other extensions of the SM which are characterized by a
complicated Higgs sector and {\em CP} violating sources in the SUSY breaking
terms of the scalar potential. The development of a charge breaking minimum
during the cooling down of the Universe may lead, in this case too, to a
large net charge asymmetry restricting the allowed parameter space. Besides
this, one has to pay special attention to the appearance of other kinds of
minima, such as color breaking, which develop at finite temperatures.
These may impose further
constraints on supersymmetry parameters and affect
the phenomenological predictions. Such a study is under
consideration and the results will appear elsewhere.

It is worth mentioning that the possible two stage transition, studied in
this paper, may be a feature shared by other multiscalar potentials
deserving special attention.
In GUT theories
the breaking of $U(1)_{em}$ symmetry at the unification scale offers a
working mechanism for the resolution of the monopole problem as proposed by
Langacker and Pi \cite{lang}.

\vspace{1.5cm} 
\noindent 
{\bf Acknowledgments}

\noindent 
We would like to thank T.W.B. Kibble and N. Tetradis for helpful
discussions. Our thanks are also due to A. Pilaftsis for useful comments and
suggestions. A.B.L. acknowledges support from ERBFMRXCT--960090.

\newpage 
\begin{table}[t]
\caption{Sample of allowed parameters respecting
all theoretical constraints
and experimental bounds on Higgs boson masses (see main text).}
\label{param}
\end{table}

\begin{center}
\begin{tabular}[t]{||c|c|c|c|c||}
\hline\hline
$\lambda _{1}/g^{2}$ & $\lambda _{2}/g^{2}$ & $\lambda _{3}/g^{2}$ & $%
\lambda _{4}/g^{2}$ & $\lambda _{5}/g^{2}$  \\ \hline\hline
1.0 & 1.0 & \hspace{1ex}1.0 & --1.0               & --1.0 \\ \hline
1.0 & 1.0 & --0.1           & --1.0               & --0.1 \\ \hline
1.0 & 1.0 & \hspace{1ex}0.1 & --1.0               & --0.1 \\ \hline
1.0 & 1.0 & \hspace{1ex}1.0 & --1.0               & --0.1 \\ \hline
1.0 & 1.0 & --0.1           & \hspace{1ex}0.1     & --1.0 \\ \hline
1.0 & 1.0 & \hspace{1ex}0.1 & \hspace{1ex}0.1     & --1.0 \\ \hline
1.0 & 1.0 & \hspace{1ex}1.0 & \hspace{1ex}0.1     & --1.0 \\ \hline\hline
\end{tabular}
\end{center}

\newpage 
\newpage 

\end{document}